\documentclass[12pt]{article}
\usepackage{amssymb}
\def\be{\begin{equation}}
\def\ee{\end{equation}}
\def\ba{\begin{eqnarray}}
\def\ea{\end{eqnarray}}
\def\>{\rangle}
\def\<{\langle}
\def\n{\nonumber}
\def\<{\langle}
\def\>{\rangle}
\def\sc{\scriptsize}
\begin{document}
\begin{center}
{\Large\bf Delayed-choice measurement and temporal nonlocality}\\
\medskip
Ilki~Kim\hspace{.5mm}$^{1}$ and G\"{u}nter~Mahler\hspace{.5mm}$^{2}$\\
\vspace*{0.1cm}
$^{1}$Center for Nano Science and Technology,
      University of Notre Dame\\
      200 Cushing Hall, Notre Dame, IN 46556, USA\\
$^{2}$Institut f\"ur Theoretische Physik~I\,, Universit\"{a}t Stuttgart\\
Pfaffenwaldring 57, 70550 Stuttgart, Germany\\
phone: ++1-219-631-3271,\, fax: ++1-219-631-4393\\
e-mail: ikim@nd.edu
\end{center}
%
%%%%%%%%%%%%%%%%%%%%%%%%%%%%%%%%%%%%%%%%%%%%%%%%%%%%%%%%%%%%%%%%%%%%%%%%%%%%%
%
\begin{abstract}
We study for a composite quantum system with a quantum Turing architecture 
the temporal non-locality of quantum mechanics by using the temporal Bell 
inequality, which will be derived for a discretized network dynamics 
by identifying the subsystem indices with (discrete) 
parameter time. However, the direct ``observation'' of the quantum system will 
lead to no violation of the temporal Bell inequality and to consistent 
histories of any subsystem. Its violation can be demonstrated, though, for a 
delayed-choice measurement.
\end{abstract}
\vspace*{0.7cm}

\begin{figure}[htbp]% fig 1
\refstepcounter{figure}\label{qtm}
\vspace*{15cm}
\begin{center}
\includegraphics{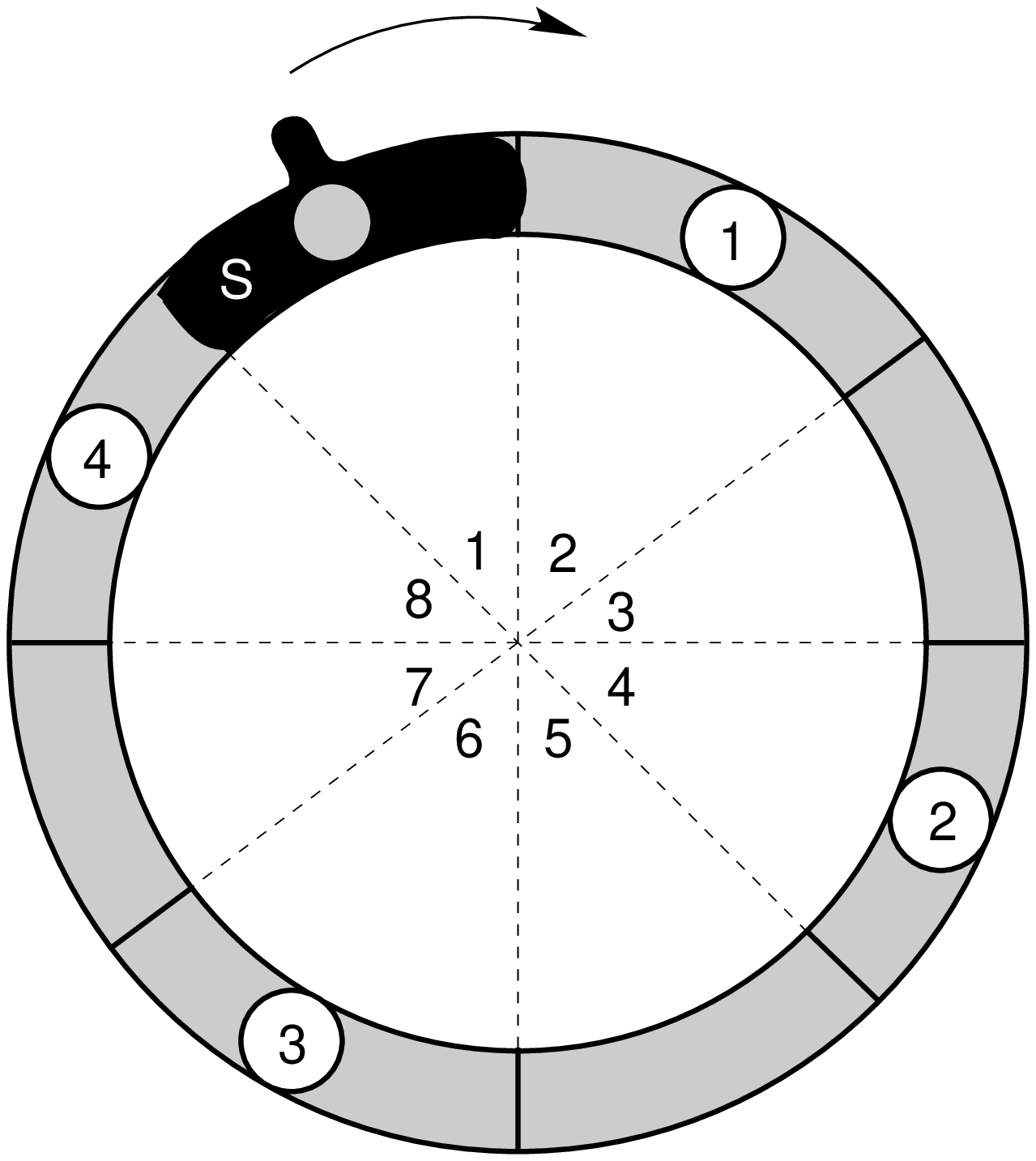}
\end{center}
\end{figure}

It has been realized from the beginning era of the quantum mechanics that 
the time can be described in quantum mechanics not by a self-adjoint operator 
but as a classical parameter in the state change defined 
by a unitary operator $\hat{U}(t)$\,. 
If the state change could be defined without 
consideration of abstract wavefunctions and their evolutions, the relationship 
between the state change and (classical) parameter time would become more 
illustrative. In this case we might ask whether the 
quantum-mechanical state change 
is local on the (classical) time axis. The state change could formally be 
described completely by a two-time correlation function, which refers to 
two points on the time axis. But this is not enough, because these 
correlations can be regarded, from their non-classical aspects, not as 
fingerprint of consistent histories \cite{OMN92} related to the temporal 
locality \cite{PAZ93}. We will show this for a composite quantum system 
(``quantum network'' \cite{MAH98}) with a quantum Turing architecture 
(see Fig.~\ref{qtm}) \cite{KIM99,KIM00} utilizing a structure of its 
internal correlations which emerge from a sequence of modular unitary 
transformations $\left\{\hat{U}_{\mu}\right\}_{\mu\in{\mathbf{N}}}$\,. 
In this way we can build a net of these correlations. 
By identifying the classical subsystem indices $\mu$\, with the discrete 
``time'', we will implement a dynamics for which the operator 
$\hat{U}_{2\mu}$\, generates at the time $2\mu$\, the correlation between a 
certain reference 
subsystem $S$\, and the subsystem $\mu$\,. Then the state of $S$\, at the 
time $2\mu$\, can be stored in the subsystem $\mu$ (``memory spins'') by 
means of the respective correlation, e.g. 
$K_{zz}^{(S,\mu)} = \left\<\hat{\sigma}_z^{(S)} \otimes 
\hat{\sigma}_z^{(\mu)}\right\>$\,, generated by a non-invasive measurement 
on $(S,\mu)$\, (e.g. CNOT-operation). The delayed-measurement of the 
correlation $K_{zz}^{(\mu_1,\mu_2)} = \left\<\hat{\sigma}_z^{(\mu_1)} \otimes 
\hat{\sigma}_z^{(\mu_2)}\right\>$\, then describes exactly the two-time 
correlation $C^{(S)}(t_1=2\mu_1;t_2=2\mu_2)$\, of $S$\, by 
reinterpreting the correlation between the subsystems $\mu_1$\, and $\mu_2$\, 
as the two-time correlations for $S$\, \cite{PAZ93}. 
This identification of the subsystem index to the time index enables us to 
construct the temporal B{\sc{ELL}} inequality, based on the standard 
B{\sc{ELL}} inequality with respect to subsystems $\mu$\,, in order to 
test the non-locality of the quantum-mechanical time evolution. 
While the temporal B{\sc{ELL}} inequality cannot be 
violated by direct measurement, it will be shown that the inequality can be 
violated by post-selection (temporal which-path 
sorting) of the states of the subsystems $\mu$\, \cite{KIM00}. One may thus 
conclude that the temporal locality and 
non-locality are complementary to each other just like in the spatial case.

The quantum network to be considered here will be composed of $M + 1$\, 
physically different $j=1/2$-spins $\nu=S, 1,2, \cdots\, M$\, (here $M=4$\,). 
The respective spin states are $\left|p^{(\nu)}\right>, p=-1, 1$\,. The 
corresponding product basis is $\left|p^{(S)}\, q^{(1)}\, r^{(2)}\, s^{(3)}\, 
t^{(4)}\right>$\,. We can describe the total spin dynamics by using the 
$SU(2)$-algebra and cluster operators 
\begin{eqnarray}
\hat{K}_{jklmn} &=& 
\hat{\sigma}_j^{(S)} \otimes \hat{\sigma}_k^{(1)} \otimes 
\hat{\sigma}_l^{(2)} \otimes \hat{\sigma}_m^{(3)} \otimes 
\hat{\sigma}_n^{(4)}\;,\n
\end{eqnarray}
where  $\hat{\sigma}_j,\, j=x,y,z$\, are the Pauli matrices. 
Let the initial network state be 
$\left|\psi(0)\right\> = \left|-1^{(S)}\right\> \left|-1^{(1)}\right\> 
\left|-1^{(2)}\right\> \left|-1^{(3)}\right\> \left|-1^{(4)}\right\>$\,. 
In the first step we apply the local transformation 
$\hat{U}^{(S)}\left(\alpha\right) = 
\exp\left\{-\frac{i}{2}\,\alpha\,\hat{\sigma}_x^{(S)}\right\}$\, 
with a phase $\alpha$\, and in the second 
step we execute the CNOT-operation on $(S,1)$\, in order to generate the 
quantum correlation ($=$\, entanglement) between $S$\, and $\mu$\, 
(only if $p(S)=-1$\,, the spin $\mu$\, will flip). 
We thus get the strict anticorrelation between $S$\, and $\mu$\,, 
$\hat{K}_{zz}^{(S,1)} = -1$\,. In the third step we again apply 
the local rotation $\hat{U}^{(S)}\left(\alpha\right)$\, 
on $S$\,, which follows the CNOT on $(S,2)$\,. 
In this way the spins $\mu=1,2,3,4$\, will ``keep'', as memories, the states 
of $S$ at the correspondng steps. Finally, we have at the $8$th step 
\begin{eqnarray}
&K_{zz}^{(1,2)}\, =\, K_{zz}^{(2,3)}\, =\, K_{zz}^{(3,4)}\, =\, 
\cos \alpha\;,\;\; K_{zz}^{(1,4)}\, =\, (\cos \alpha)^3\;.&\n
\end{eqnarray}
Fig.~\ref{history}\, 
shows the time evolution of spin $S$ and the memory states at all 
steps. Via this built-in logic we interpret the correlations 
$K_{zz}^{(j,k)}$\, as the two-time correlation functions 
$C^{(S)}(2j, 2k)$\, between steps $2j$\, and $2k$\,. 
The temporal Bell inequality reads
\begin{eqnarray}
&&\left|K_{zz}^{(1,2)}\, +\, 
K_{zz}^{\left(2,3\right)}\, +\, K_{zz}^{\left(3,4\right)}\, -\, 
K_{zz}^{\left(1,4\right)}\right|\; \Rightarrow\n\\
&&|C^{\left(S\right)}\left(2,4\right)\, +\, 
C^{\left(S\right)}\left(4,6\right)\, +\, 
C^{\left(S\right)}\left(6,8\right)\, -\, 
C^{\left(S\right)}\left(2,8\right)|\; \leq\; 2\n\\
&&\left|\cos{\alpha}\, +\, \cos{\alpha}\, +\, \cos{\alpha}\, -\, 
\cos\left(\alpha + \alpha + \alpha\right)\right|\; \leq\; 2\;.\n
\end{eqnarray}
However, this inequality cannot be violated \cite{KIM00}: each possible 
history of $S$\, can be considered an ``element of reality'' because 
subsequent measurements of the memories $\mu$\, 
would project the subsystem $S$\, into an individual temporal trajectory of 
$S$\, describing a specific history of $S$\,. The perturbation by the 
non-invasive measurement at the step $2$\, and $4$\, leads to a form of the 
correlation which differs from the quantum mechanical result 
$C^{(S)}(2,8)=\cos(3 \alpha)$\, to be expected for the isolated 
unitary dynamics of $S$\,. Therefore the temporal 
non-locality could be verified experimentally only by ``incompatible'' 
measurements of these two-time correlations. In Fig.~\ref{history}\, 
we see that a set of $4$\, measured values of spin 
components $\left(\sigma_z^{(1)}, 
\sigma_z^{(2)}, \sigma_z^{(3)}, \sigma_z^{(4)}\right)$\,, 
$\sigma_z^{(\mu)}=\pm 1$\,, clearly allocates a specific history 
$C_j\,,\, j=1,2, \cdots\, 2^4$\,. Similarly we can also build another 
consistent history of $S$\, by considering $\sigma_x^{(\mu)} = \pm 1$\, 
$\left(\mbox{not}\;\; \sigma_z^{(\mu)} = \pm 1\right)$\, 
with keeping an event of $S$ at each step. 
In this case we get a different kind 
of history $C_j'\, \left(\sigma_x^{(1)}, \sigma_x^{(2)}, \sigma_x^{(3)}, 
\sigma_x^{(4)}\right)$\,. Because of the incompatibility of 
$\hat{\sigma}_x^{(\mu)}$\, and $\hat{\sigma}_z^{(\mu)}$\, 
$\left(\,\left[\hat{\sigma}_x^{(\mu)} , \hat{\sigma}_z^{(\mu)}\right] 
\ne 0\right)$\, we cannot have the above two kinds of histories 
simultaneously. However, based on 
\begin{eqnarray}
&\left|-1_x^{\left(\mu\right)}\right\>\; =\; \left(
\left|-1_z^{\left(\mu\right)}\right\>\, -\, 
\left|1_z^{\left(\mu\right)}\right\>\right)/
\sqrt{2}\;,\; \left|1_x^{\left(\mu\right)}\right\>\; =\; \left(
\left|-1_z^{\left(\mu\right)}\right\>\, +\, 
\left|1_z^{\left(\mu\right)}\right)\right\>/\sqrt{2}&\n
\end{eqnarray}
each history $C_j'$\, can be rewritten as a coherent superposition of all 
$2^4$\, histories $C_j\,,\, j=1,2, \cdots\, 2^4$\,. Then it follows for each 
history $C_j'$\, that there is no projection 
of $\hat{\sigma}_z$\, available on the $\sigma_z$-axis but only local 
rotations of spin $S$\, with respect to the $\sigma_x$-axis. Therefore all 
histories $C_j'$\, can be distinguished by total rotating angles 
of $S$\,: e.g. at the second step we have two histories for each case 
$C_j \left(C_j'\right)\,,\, j=1,2$\, with the following network state
\begin{eqnarray}
&|\psi\>\, =\, \cos\left(\alpha/2\right)\, 
\underbrace{\left|-1_z^{\left(S\right)}\right\>\, 
\left|1_z^{\left(\mu\right)}\right\>}_{C_1}\, -\, i\, 
\sin\left(\alpha/2\right)\, \underbrace{\left|1_z^{\left(S\right)}\right\>\, 
\left|-1_z^{\left(\mu\right)}\right\>}_{C_2}&\n\\
&=\, \frac{1}{\sqrt{2}}\, 
\left(\exp\left\{-\frac{i}{2}\,\alpha\,\hat{\sigma}_x^{(S)}\right\}
\underbrace{\left|-1_z^{\left(S\right)}\right\> \otimes 
\left|1_x^{\left(\mu\right)}\right\>}_{C_1'}\, -\, 
\exp\left\{-\frac{i}{2}(-\alpha)\,
\hat{\sigma}_x^{(S)}\right\}\underbrace{\left|-1_z^{\left(S\right)}\right\> 
\otimes \left|-1_x^{\left(\mu\right)}\right\>}_{C_2'}\right)&\,.\n
\end{eqnarray}
Such a ``weak'' non-invasive measurement (leading to an ignorance of the 
temporal which-path information) enables us to reconstruct a unitary 
(coherent) dynamics of $S$\, (already having temporal which-path information 
by $\sigma_z^{(\mu)} = \pm 1$)\, 
as if the non-invasive measurement had not been 
applied to $S$\,. While both which-path markings by 
$\sigma_x^{(\mu)}=\pm 1$\, and $\sigma_z^{(\mu)}=\pm 1$\, generate in each 
case consistent internal histories, each history $C_j'\; (C_j)$\, has but no 
temporal which-path marking with respect to 
$\sigma_z^{(\mu)}\, \left(\sigma_x^{(\mu)}\right) = \pm 1$\,. 
From this incompatibility of these markings a delayed-measurement of 
$\sigma_x^{(\mu)}\, \left(\sigma_z^{(\mu)}\right) = \pm 1$\, at the end step 
of the dynamics leads to the coherent superposition of 
$\sigma_z^{(\mu)}\, \left(\sigma_x^{(\mu)}\right) = \pm 1$\,. 
Therefore the measurement of $\sigma_x^{(\mu)}$\, erases the temporal 
which-path information of $S$\, marked by $\sigma_z^{(\mu)}$\, by considering 
the ignorance of the ``inter''-consistent histories between $C_j$\, and 
$C_j'$\,. Then we do not have consistent histories any more. This back-action 
on the state in the past already indicates the temporally non-local aspect of 
quantum mechanics. Now we consider the correlation $K_{zz}^{(1,4)}$\, in the 
B{\sc{ELL}} inequality. Before applying the delayed-measurement 
of the correlation functions we measure 
$\hat{\sigma}_x^{(2)}\,,\, \hat{\sigma}_x^{(3)}$\, 
and select the corresponding network states e.g. those with 
$\sigma_x^{(2)}=\sigma_x^{(3)}=+1$\,. After doing this we have the 
superposition of the four histories with 
\begin{eqnarray}
\sigma_z^{(2)}\; =\; \sigma_z^{(3)}\; =\; +1 &,&
\sigma_z^{(2)}\; =\; -\sigma_z^{(3)}\; =\; +1\n\\
\sigma_z^{(2)}\; =\; \sigma_z^{(3)}\; =\; -1 &,&
\sigma_z^{(2)}\; =\; -\sigma_z^{(3)}\; =\; -1\n
\end{eqnarray}
and the selected nework state $|\psi_s\>$\, reads as 
\begin{eqnarray}
\left|\psi_s\right\>\; &=&\; \cos\left(\alpha/2\right) 
\cdot \cos\left(3\alpha/2\right)\, 
\left|1_z^{(4)}\,1_x^{(3)}\,1_x^{(2)}\,1_z^{(1)}\right\> 
\left|-1^{(S)}\right\>\; +\n\\
&& -i\, \cos\left(\alpha/2\right) \cdot \sin\left(3\alpha/2\right)\, 
\left|-1_z^{(4)}\,1_x^{(3)}\,1_x^{(2)}\,1_z^{(1)}\right\> 
\left|1^{(S)}\right\>\; -\n\\
&& \sin\left(\alpha/2\right) \cdot \sin\left(3\alpha/2\right)\, 
\left|1_z^{(4)}\,1_x^{(3)}\,1_x^{(2)}\,(-1)_x^{(1)}\right\> 
\left|-1^{(S)}\right\>\; +\n\\
&& -i\, \sin\left(\alpha/2\right) \cdot \cos\left(3\alpha/2\right)\, 
\left|-1_z^{(4)}\,1_x^{(3)}\,1_x^{(2)}\,(-1)_z^{(1)}\right\> 
\left|-1^{(S)}\right\>\;.\n
\end{eqnarray}
As a result $4$\, histories $C_j$\, with 
$\sigma_z^{(\mu)}=\pm 1\,,\, \mu=2,3$\, 
have been erased (temporal quantum erasure) \cite{SCU91,KIM00}. 
Now we execute the delayed-measurement of $K_{zz}^{(1,4)}$\, and finally get 
the correct quantum mechanical form of correlation function 
$K_{zz}^{(1,4)} = \cos(3\alpha)$\, as if there had been no influence 
by the memories $2, 3$\, on the unitary dynamics of $S$\,. The same holds 
for other post-selected states, i.e. for other $\sigma_x^{(2)}$\, and 
$\sigma_x^{(3)}$\, results. In this way the violation of the temporal 
B{\sc{ELL}} inequality becomes measurable, e.g. with $\alpha = \pi/4$\, 
\begin{eqnarray}
&&\left|\cos{\alpha}\, +\, \cos{\alpha}\, +\, \cos{\alpha}\, -\, 
\cos\left(3\alpha\right)\right|\, =\, 
\left|\, 3\, \cos\left(\pi/4\right)\, -\, 
\cos\left(3\pi/4\right)\, \right|\, =\, 2 \sqrt{2}\; \nleq\; 2\;.\n
\end{eqnarray}
This violation shows explicitly that the quantum mechanical state change 
$\hat{U}(t)$\, (before measurement) is, in general, temporally non-local 
and we therefore cannot expect consistent histories for a quantum-mechanical 
system under unitary evolution.

We thank J.~Gemmer, A.~Otte, M. Stollsteimer, F. Tonner and T.~Wahl 
for fruitful discussions.
%
%%%%%%%%%%%%%%%%%%%%%%%%%%%%%%%%%%%%%%%%%%%%%%%%%%%%%%%%%%%%%%%%%%%%%%%%%%%%%
% References
%%%%%%%%%%%%%%%%%%%%%%%%%%%%%%%%%%%%%%%%%%%%%%%%%%%%%%%%%%%%%%%%%%%%%%%%%%%%%
%

\noindent{Figure~\ref{qtm}: 
Quantum network with a quantum Turing architecture: a reference 
subsystem~$S$\, (Turing head) and memory 
spins~$\mu = 1,2,3,4$\,; step number (time) $= 1,2, \cdots\, 8$ 
($=$ position of $S$)}\,.
\vspace*{1cm}

\noindent{Figure~\ref{history}: Alternative histories of $S$\,; 
$0 \equiv \left|-1\right\>$\, and $1 \equiv \left|1\right\>$\,.}
\newpage
\newpage
\begin{figure}[htbp]% fig 2
\hspace*{1cm}
\refstepcounter{figure}\label{history}
\includegraphics{figure2.ps}
\end{figure}
%%%%%%%%%%%%%%%%%%%%%%%%%%%%%%%%%%%%%%%%%%%%%%%%%%%%%%%%%%%%%%%%%%%%%%%%%%%%%
\end{document}